\documentclass[preprint,12pt]{elsarticle}
\usepackage[ruled,vlined]{algorithm2e}
\usepackage{multirow}
\usepackage{array}
\usepackage{amsmath}
\usepackage{enumerate}
\usepackage{extsizes}
\usepackage[version=3]{mhchem}
\usepackage[left=1.5cm, right=1.5cm, top=1.785cm, bottom=2.0cm]{geometry}
\usepackage{balance}
\usepackage{times,mathptmx}
\usepackage{sectsty}
\usepackage{graphicx}
\usepackage{lastpage}
\usepackage[format=plain,justification=raggedright,singlelinecheck=false,font={stretch=1.125,small,sf},labelfont=bf,labelsep=space]{caption}
\usepackage{float}
\usepackage{fancyhdr}
\usepackage{fnpos}
\usepackage[english]{babel}
\usepackage{array}
\usepackage{droidsans}
\usepackage{charter}
\usepackage[T1]{fontenc}
\usepackage[usenames,dvipsnames]{xcolor}
\usepackage{setspace}
\usepackage[compact]{titlesec}

\usepackage{epstopdf}%This line makes .eps figures into .pdf - please comment out if not required.

\usepackage{amssymb}
\usepackage{amsmath}
\usepackage{lineno}
\usepackage{caption}
\usepackage{subcaption}
\usepackage{setspace} 
\journal{}

\begin{document}

\begin{frontmatter}
\title{Disease gene prioritization using network topological analysis from a sequence based human functional linkage network}
\author{A. Jalilvand\textit{$^{a}$}, B. Akbari,\textit{$^{a\ast}$}, F. Zare Mirakabad\textit{$^{b}$} and \textit F. Ghaderi{$^{a}$} }
\address{a: Department of Electronic and computer engineering,Tarbiat Modares University, Tehran, Iran\\
b: Department of Mathematics and Computer Science, Amirkabir University of
Technology, Tehran, Iran\\
Correspondence: b.akbari@modares.ac.ir
}
\begin{abstract}
Sequencing large number of candidate disease genes which cause diseases in order to identify the relationship between them is an expensive and time-consuming task. To handle these challenges, different computational approaches have been developed. Based on the observation that genes associated with similar diseases have a higher likelihood of interaction, a large class of these approaches relay on analyzing the topological properties of biological networks.
However, the incomplete and noisy nature of biological networks is known as an important challenge in these approaches. In this paper, we propose a two-step framework for disease gene prioritization: (1) construction of a reliable human FLN using sequence information and machine learning techniques, (2) prioritizing the disease gene relations based on the constructed FLN. On our framework, unlike other FLN based frameworks that using FLNs based on integration of various low quality biological data, the sequence of proteins is used as the comprehensive data to construct a reliable initial network. In addition, the physicochemical properties of amino-acids are employed to describe the functionality of proteins. All in all, the proposed approach is evaluated and the results indicate the high efficiency and validity of the FLN in disease gene prioritization.
\end{abstract}
\end{frontmatter}
%\linenumbers

\section{Introduction}
\label{Sec:IntroSec}
Genetic diseases are caused by malfunctioning of a single or multiple genes. According to this definition, these diseases can be categorized into single gene diseases and polygenic diseases (complex diseases)\cite{tenesa2013heritability}. Polygenic disease such as diabetes, alzheimer, and cancers are more common compared with single-gene diseases. Therefore, prioritizing complex diseases gene is a fundamental goal in computational biomedicine.

Traditional linkage analysis and association studies have been successful in locating loci and identifying a large number of candidate genes that probably are associated with diseases \cite{lei2019predicting,risch1996future}.Since sequencing all the identified candidate genes is a costly and time-consuming task, the computational approaches have been developed to deal with this challenge. These approaches can be divided into two general categories:
Firstly, a number of machine-learning methods have been introduced to compute the similarity scores among candidate genes with the known disease genes (seed genes). These methods are based on the observation that the genes of the same disease probably share similar biological properties. In this category, different biological data such as gene sequence data, gene expression profile\cite{Vandamexpression} , functional annotation data \cite{Tiffin,eguchi2018integrative}, and integrating multiple data sources\cite{aerts2006gene,eguchi2018integrative} are applied to generate the feature vectors. So according to these feature-vectors, the candidate genes are scoring as disease genes. The main issue with them is the incomplete and noisy nature of biological data.
Secondly, some of the computational approaches use different scoring strategies to prioritize candidate genes according to seed genes in biological networks\cite{wang2011integration, liu2015prioritization}. These network-based methods are usually introduced according to the underlying assumption of "guilt-by-association" which is a title with this definition; the genes which cause same diseases or exist in a certain pathway, should be physically or functionally similar together\cite{lei2019predicting,findlay2018guilt,barabasi2011network}.
The Protein-Protein Interaction (PPI) is one of the famous biological networks that has been used to detect protein complexes\cite{he2018evolutionary}, identify essential proteins or genes\cite{bbz017}, and predict protein functions\cite{herzog2018detection,schaefer2012evolution,wong2015imp}. Therefore, some of the computational methods use the PPI network for disease genes prioritization based on the distance between the candidate genes and the seed genes \cite{wang2011integration, liu2015prioritization}. Oti et al.\cite{oti2006predicting} used direct neighbor-based method to measure the distance between the candidate genes and the seed genes. Hsu et al.\cite{hsu2011prioritizing} proposed a nearest-neighbor-based method to prioritize candidate disease genes. They used the interconnectedness (ICN) measure to prioritize candidate genes by calculating the closeness between them and seed genes in the PPI network.

Recent studies in the systems biology show that the similar diseases are more caused by functionally related genes than physically interacting genes. Therefore, the methods based on PPI networks with the local distance measures such as neighbor methods are unable to reach satisfying results in prioritization disease genes. There are two ways to tackle this challenge, 1) Some researchers have used the global distance measures such as random walk and shortest path to measure the closeness between two candidate and seed genes. Zhu et al. \cite{zhu2012vertex} introduced a Vertex Similarity method based on shortest paths to prioritize candidate genes. Li et al.\cite{li2014prioritization} introduced two novel shortest path methods to prioritize candidate disease genes in PPI networks. 2) In the second class of approaches, according to the fact that applying a comprehensive initial network (such as FLN) is an essential step in disease gene prioritization framework, researchers have developed methods to construct such a suitable FLN\cite{Wang2011, manimaran2009prediction,apolloni2011learning}.
FLNs are well-defined data structures, which are used to identify disease genes\cite{apolloni2011learning}. An FLN is a graph in which the nodes represent genes or corresponding proteins and the edges denote functional associations between them. In other words, two proteins are connected in an FLN if some experimental or computational methods indicate that they share the same functionality\cite{jalilvand2018s}.

Many research works have been focused on network reconstruction methods, which are continued to integrate different data sources to construct the biological network\cite{you2010using,lei2012assessing,Franke2006,Linghu2009,Wang2014,Wu2010}. 
An FLN is constructed by integrating various types of biological data \cite{Franke2006}. The authors employed PPI, microarray co-expression, and GO(Gene Ontology) along with applying a Bayesian approach to predict gene pairs that participate in the same GO biological process. Similarly, in \cite{kohler2008walking} multiple biological data sources using random walk algorithm prioritized disease genes.
In \cite{Linghu2008}, a six-dimensional feature vector extracted as the six biological data sources has been proposed. Then, multiple machine learning methods such as Support Vector Machine (SVM), Linear Discriminant Analysis(LDA), Naive Bayes, and Neural Network were applied to construct a reliable FLN. 
A human FLN by integrating 16 biological data features from 6 model organisms has been constructed by \cite{Linghu2009}. Afterwards, they use a Naive Bayes classifier to predict functional linkage between genes. Wang et al. \cite{Wang2014} built an FLN of mitochondrial proteins by integrating biological features such as genomic context, gene expression profiles, metabolic pathways and PPI network.  
A recent and interesting survey on functional linkage network construction methods is provided in \cite{Linghu2013}.
Although integration methods might be acceptable in social networks and other non-biological networks, due to low quality of biological data sources they are faced with major challenges in biological networks.

A number of methods are introduced in the literature to overcome this issue, in which information of amino acid sequences is applied to predict links in biological networks\cite{singh2018evolutionary,shen2007predicting,guo2008using,xia2010predicting,yousef2013novel,you2013prediction,mei2014adaboost}. 
These methods can be categorized into two classes, alignment-based\cite{borozan2015integrating,Yushuang2016,otu2003new} and alignment-free methods. The alignment-free methods have been proposed to tackle challenges such as inaccuracy on the inversion, translocation at substring level, and diverse sequences with the same functionally or unequal lengths \cite{zhang2011adaptive,xia2010predicting,yousef2013novel,yang2010prediction,shen2007predicting}. Some of these methods use physicochemical properties of amino acids to enrich extracted feature vectors\cite{xia2010predicting,yousef2013novel,huang2016sequence}.
A computational approach based on compressed sensing theory is introduced in \cite{zhang2011adaptive} to predict yeast PPI. They have used Auto Covariance (AC) method\cite{guo2008using} with 7 physicochemical properties to extract the features. A computational model has been proposed in  \cite{huang2016sequence} to predict PPIs by combining a global encoding representation of sequences and a weighted sparse representation-based classifier.
In \cite{xia2010predicting,you2013prediction,yousef2013novel}, six feature descriptor methods have been applied in an ensemble approach, including Auto Covariance (AC)\cite{guo2008using}, Geary Autocorrelation (GA)\cite{sokal2006population}, Conjoint Triad (CT)\cite{shen2007predicting}, Local Descriptor (LD)\cite{yang2010prediction}, Moran Autocorrelation(MA)\cite{xia2010sequence} and Normalized Moreau-broto Autocorrelation(NMA)\cite{feng2000prediction}. 
The obtained results show that employing data fusion and ensemble learning structures provide acceptable accuracy in PPI link prediction. 
It has been found in many researches that the information of amino acid sequences are sufficient to predict protein-protein interactions \cite{xia2010predicting,mei2014adaboost,guo2008using}. 
Although the beneficial effects of physicochemical properties of amino-acid in modeling functional relationships is obvious, there has not been any method to construct FLN based on amino acid sequences to be used in disease genes prioritization so far. 
Therefore, in a previous research called sequence-based FLN (S-FLN) we developed an efficient approach to construct the yeast FLN using sequence information of amino acids \cite{jalilvand2018s}. In this work, we propose a network-based framework to prioritize disease genes using a sequence-based FLN as an initial network.   
We evaluate the proposed framework using two different versions of OMIM disease gene datasets. The proposed framework obtained 96\% and 94.3\% AUC rates on each dataset. 
In summary, the main ideas of our paper are: 
\begin{itemize}
\item Considering the fact that human diseases result from perturbations of molecular networks, we propose a network-based computational framework for prioritizing disease-related genes to handle the aforementioned challenges.
\item We evaluate the impact of selecting appropriate initial networks in disease gene prioritization application and compared S-FLN with other simple FLN and PPI networks. 
\item Using physicochemical properties of amino acids and various descriptor methods in the proposed network-based framework, we show that the S-FLN is useful to capture various functional linkages and consequently different disease.
\end{itemize}

The remaining of this paper has been organized as follows. In section 2 the basic concepts and biological network construction approaches are briefly reviewed. In section 3 the proposed approach is introduced. The experimental results are presented in section 4. In section 5 the performance of S-FLN is discussed, and finally we conclude the paper in section 6.

\section{Problem Definition}
The aim of disease gene prioritization is to score a set of candidate genes $C_{set} = (c_1,c_2,...c_n)$ as a n-triplet where the value in the $i^{th}$ position indicates the score of the $i^{th}$-relevant gene, which are ordered according to the probability of the effect of the query gene in a specific disease processes.
In order to score the $C_{set}$, we need to calculate distance between $C_{set}$ and $S_{set}$ in an initial network such as constructed FLN $G$, as stated in definition 1.
Hence according to the influence of a reliable initial network in prioritizing disease gene process, the first step is predicting links between the protein pairs to construct the FLN. The goal of FLN construction is to build a graph $G = (V, P)$ which is build to be as similar as to the natural FLN graph $G_N = (V_p, E_p)$, as stated in definition 2 and 4.
In the second step, $C_{set}$ and $S_{set}$ are mapped on the constructed FLN $G$ as initial network. The distance of each member in $C_{set}$  with $S_{set}$ has been calculated by different distance measures. The output of proposed disease gene prioritization approach is a set of scored candidate genes $C_{S_{set}} = (c_1=1,c_2=0.8,...c_n=0)$ according to these distance measures.
\begin{description}
\item[Definition 1.] Candidate gene the set $C_{set}$ is the set of genes that has the probability to cause a Mendelian disorder, or contributing to a complex disease. Seed gene set $S_{set}$ is the set of known disease genes.
\item [Definition 2.] Natural FLN: is a graph  $G_N = (V_p, E_p)$ where $V_p = P = \{p_1,p_2,...,p_n\}$ is a set of $n$ proteins or corresponding genes and $E_p \subseteq P \times P$ is the set of functional links between the protein pairs.
\item [Definition 3.]Gold Standard: is a graph  $G_S = (V_g, E_g)$ where $V_g = G = \{g_1,g_2,...,g_m\}$ is a set of  $m \leqslant n$  genes that correspond to a protein as their product in the Natural FLN. Here \textit{n} is the number of all proteins from \textit{Definition 1}, $E_g \subseteq G \times G$ is the set of functional links between a pair of genes. 
%These links are calculated based on GO dataset.
\item [Definition 4.]Constructed FLN is a predicted graph  $G = (V, E)$ where $V = P = \{p_1,p_2,...,p_n\}$ is a set of $n$ proteins and $E \subseteq E_p$ is the set of predicted functional links between a pair of proteins. 
\end{description}

Table 1 summarizes the notation used in this paper.

\begin{table}[h!]
	\small
	\caption{\ The table of symbols and their descriptions.}
	\label{tbl:symbol}
	\begin{tabular*}{1\textwidth}{@{\extracolsep{\fill}}ll}
		\hline
		Symbol-Variable & Definition\\
		\hline
		$S_{set}$ & The set of known disease genes (Seed gene)\\
		$C_{set}$ & The set of genes that are more likely to causing a Mendelian disorder \\
		$C_{score}$ & The set of candidate genes that are scored according to probability of being a disease genes  \\
		$DG_{set}$ & The set of polygenic diseases and their genes that include $S_{set}$ and $C_{set}$ \\
		$G_N=(V_p, E_p)$  & The natural FLN graph that includes proteins as vertices $V_p$ and functional linkages as edge $E_p$\\
		$G_S=(V_g, E_g)$  & The gold standard graph that includes proteins as vertices $V_g$ and functional linkages as edge $E_g$\\
		$G=(V, E)$  & The constructed graph include proteins as vertices $V$ and predicted functional linkages as edge $E$\\
		$S_s$ & Similarity weight between a pair of proteins \\
		%P  & protein set\\
		%$GO_{p_i}$ & GO terms for the $i\ th$ protein \\
		$t_{p_i}^{l_i}$ & The $l_i\ th$ term for the $p_i\ th$ protein \\
		$S_i$ & Disease association score of $i th$ candidate gene\\ 
		$N={1,...,n}$ & Neighbor set of each node  \\
		\hline
	\end{tabular*}
\end{table}

\begin{figure*}[t!]
	\centering
	\includegraphics[width=\textwidth]{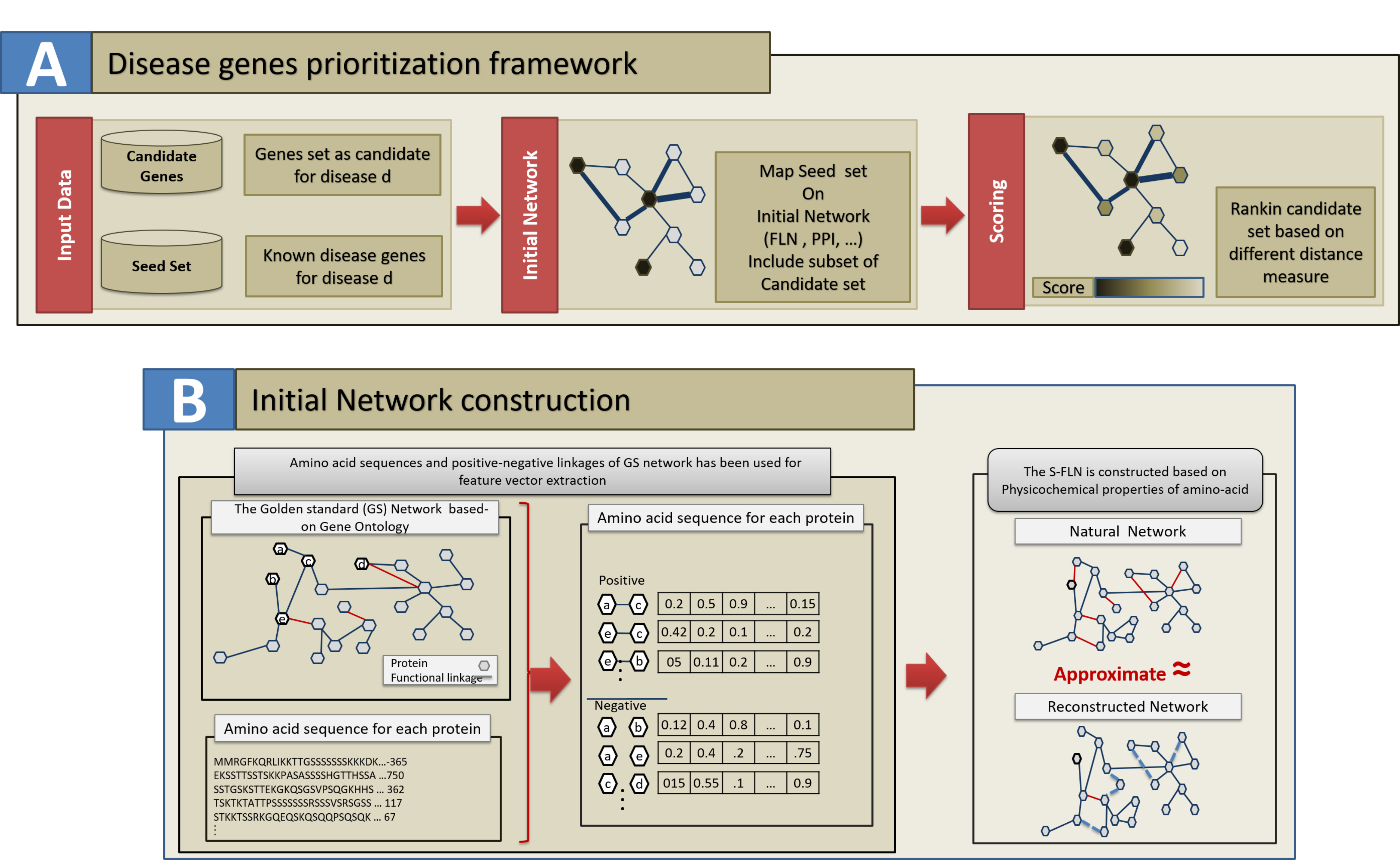}
	\caption{A: A global network-based gene prioritization framework B: The sequence-based FLN construction as initial network that is essential step in the network-based gene prioritization framework. 
	}
	\label{fgr:example2appproch}
\end{figure*}

\section{The Proposed Method}
The main goal of this research is to improve the accuracy of disease gene prioritization task in order to recognize new disease genes. Therefore, we need to obtain an approximated network that is close to the natural functional network as initial backbone network. Figure \ref{fgr:example2appproch} shows an overview of the disease gene prioritization framework. For this purpose, 
in this section we introduce a hierarchical framework consisting of two step as follows: 
\begin{itemize}
	\item Initial network construction: at first we construct the S-FLN using physiochemical properties of amino-acids as a complete and reliable biological information (see our previous work for more details \cite{jalilvand2018s}).
	\item Disease genes Prioritization: at the second step we map seed genes (known as disease genes) and candidate genes on S-FLN, then, the network topological measures are used to rank candidate genes on the S-FLN. 
\end{itemize}
%
%As mentioned in problem definition, supervised link prediction is the the main task in S-FLN. 
%Firstly, by using the gold standard graph  $Gs = (G, E)$, i.e. definition 2, an annotated dataset $GSs = \bigcup \{GSP,GSN\}$ is created. The gold standard set $GSs$ would include both the gold standard positive (GSP) and negative (GSN) sets. These sets are used as the train and test data. 
%Then, a set of feature vectors $F=\{f_1 , . . ., f_n\}$ for each protein pair in $GSs = \bigcup \{GSP,GSN\}$ is extracted, 
%where the protein sequence information is used as input data. 
%Finally, a supervised learning algorithm would be employed on the training set to obtain the link predictor.
%In following, these three steps are described in more details.

\subsection{Constructing S-FLN}
In this section, we introduce the proposed S-FLN using the gold standard and physicochemical properties of amino-acids as additional information. See our previous paper \cite{jalilvand2018s} for more detail of S-FLN construction processes, the S-FLN has three main steps. 

\textbf{Gold standard (GS) construction:} Since we employ a supervised learning approach, a data set is required to train and test the learning method. Therefore the first step is to obtain a primary network with the highest similarity to the natural FLN. In order to reconstruct the aforementioned primary network, we use Gene Ontology (GO) as a gold standard network which comprises the training and test sets. For this purpose, a context based similarity measure is employed to capture both semantic and topological properties of the GO graph. 
To do so, we develop a two-stage similarity measure algorithm. In the first stage of the algorithm, we compute GO-term semantic similarities by aggregating information content (AIC) as an IC-based method \cite{song2014measure}. 

\textbf{Feature extraction:} as mentioned above, the GS network includes negative and positive pairs. 
In this step, we need an efficient feature extraction method to discriminate the positive pairs form the negative ones. For this purpose, seven descriptor methods are employed to extract an alignment-free feature vector for each pair of proteins, which are enriched with physicochemical properties of the amino-acids sequences. The normalized form of physicochemical properties are shown in Table 2.
Therefore, a feature vector is extracted for each protein in a pair. Afterwards, the two feature vectors would be concatenated to compose a unique feature vector.
Finally, the set of feature vectors are obtained to fed into different learners in the next step.

\begin{table*}[t!]
	\label{tbl:pysicopro}
	\small
	\caption{Normalized values of 12 physicochemical properties of amino-acids.  These properties include hydrophobicity (HY-PHOB)\cite{sweet1983correlation}, hydrophilicity (HY-PHIL)\cite{hopp1981prediction}, polarity (POL)\cite{grantham1974amino}, polarizability (POL2)\cite{charton1982structural}, solvation free energy (SFE)\cite{eisenberg1986solvation}, graph shape index (GSI)\cite{fauchere1988amino}, transfer free energy (TFE)\cite{janin1979surface}, amino acid composition (AAC)\cite{grantham1974amino}, CC in regression analysis (CC)\cite{prabhakaran1982shape}, residue accessible surface area in tripeptide (RAS)\cite{chothia1976nature}, partition coefficient (PC)\cite{ garel1973coefficients} and entropy of formation (EOF)\cite{hutchens1970heat}}
	\label{tbl:amino-properties}
	\begin{tabular*}{1.0\textwidth}{@{\extracolsep{\fill}}l|lllllllllllll}
		\hline
		& HY-PHOB & HY-PHIL & POL & POL2 & SFE & GSI & TFE & AAC & CC & RAS & PC & EOF \\
		\hline
		A & 0.281 & 0.453 & 0.395 & 0.112 & 0.589 & 0.305 & 0.777 & 0 & 0.942 & 0.222 & 0.033 & 0.124\\
		C & 0.458 & 0.375 & 0.074 & 0.312 & 0.527 & 0.422 & 1 & 1 & 0 & 0.333 & 0.033 & 0.431\\
		D & 0 & 1 & 1 & 0.256 & 0.191 & 0.381 & 0.444 & 0.501 & 0.82 & 0.416 & 0.021 & 0.314\\
		E & 0.027 & 1 & 0.913 & 0.369 & 0.285 & 0.372 & 0.407 & 0.334 & 0.902 & 0.638 & 0.042 & 0.447\\
		F & 1 & 0.14 & 0.037 & 0.709 & 0.936 & 0.701 & 0.851 & 0 & 0.697 & 0.75 & 0.372 & 0.36\\
		G & 0.198 & 0.531 & 0.506 & 0 & 0.446 & 0 & 0.777 & 0.269 & 0.904 & 0 & 0.014 & 0\\
		H & 0.207 & 0.453 & 0.679 & 0.562 & 0.582 & 0.713 & 0.629 & 0.21 & 0.735 & 0.666 & 0.021 & 0.537\\
		I & 0.792 & 0.25 & 0.037 & 0.454 & 0.851 & 1 & 0.925 & 0 & 0.668 & 0.555 & 0.13 & 0.494\\
		K & 0.198 & 1 & 0.79 & 0.535 & 0.325 & 0.451 & 0 & 0.12 & 0.32 & 0.694 & 0 & 0.809\\
		L & 0.783 & 0.25 & 0 & 0.454 & 0.851 & 0.618 & 0.851 & 0 & 0.617 & 0.527 & 0.162 & 0.489\\
		M & 0.721 & 0.328 & 0.098 & 0.54 & 0.957 & 0.56 & 0.814 & 0 & 0.144 & 0.611 & 0.115 & 0.35\\
		N & 0.12 & 0.562 & 0.827 & 0.327 & 0.319 & 0.381 & 0.481 & 0.483 & 0.502 & 0.472 & 0.028 & 0.375\\
		P & 0.253 & 0.531 & 0.382 & 0.32 & 0.702 & 0.637 & 0.555 & 0.141 & 0.748 & 0.388 & 0.053 & 0.244\\
		Q & 0.123 & 0.562 & 0.691 & 0.44 & 0.4 & 0.372 & 0.407 & 0.323 & 0.586 & 0.583 & 0.046 & 0.504\\
		R & 0.222 & 1 & 0.691 & 0.711 & 0 & 0.558 & 0.148 & 0.236 & 0.726 & 0.833 & 0.001 & 1\\
		S & 0.235 & 0.578 & 0.53 & 0.151 & 0.448 & 0.312 & 0.629 & 0.516 & 0.953 & 0.222 & 0.005 & 0.216\\
		T & 0.318 & 0.468 & 0.456 & 0.264 & 0.557 & 0.723 & 0.592 & 0.258 & 1 & 0.361 & 0.021 & 0.365\\
		V & 0.687 & 0.296 & 0.123 & 0.342 & 0.765 & 0.875 & 0.888 & 0 & 0.591 & 0.444 & 0.09 & 0.373\\
		W & 0.56 & 0 & 0.061 & 1 & 1 & 0.766 & 0.777 & 0.047 & 0.82 & 1 & 1 & 0.511\\
		Y & 0.922 & 0.171 & 0.16 & 0.728 & 0.787 & 0.701 & 0.518 & 0.072 & 0.515 & 0.861 & 0.208 & 0.475\\
		\hline
	\end{tabular*}
\end{table*}

\textbf{Ensemble learning:} In the third step, we apply a learning algorithm on the extracted feature sets. To do this, a stacking two-layer learning structure is proposed to construct the FLN. Firstly, a protein pair has seven individual feature vectors, thus seven predicted scores are obtained for each of them. Random Forest learning method is applied on seven distinct datasets in the first layer of our ensemble learning schema. 
The prediction scores of each protein pair are integrated as new \textit{7}-dimensional feature vectors. 
Then, in second layer these new features are fed to a multi-layer perceptron (MLP) classifier as the inputs to obtain the final score of each protein pair.

The major challenge in this step is the imbalanced data problem, that means the number of positive pairs set is much smaller than that of negative pairs set. As stated by Herrera et al. \cite{galar2012review}, bagging is one of the efficient methods to deal with the imbalanced data issue. Accordingly, we use a bagging method to balance the data-set.

\subsection{Disease Genes Prioritization} 
We introduced the S-FLN \cite{jalilvand2018s} as a comprehensive functional network in disease genes prioritize application.
% to capture association of genes in candidate set with those in seed set.
The assumption of  "guilt-by- association" is the main idea behind the network-based disease gene prioritization methods
\cite{barabasi2011network}. Therefore, the distance between candidate disease genes and seed genes is a suitable measure to score candidate disease genes. 

In this work, we applied different distance measures to obtain an accurate score for candidate disease genes and to evaluate the S-FLN performance in the application of disease genes prioritization.
The distance measures used in this work include two local network-search methods (Direct Link, Shortest Path) and the Random Walk as a global distance measure. Wang et al. have a review on distance measure methods in disease gene prioritization application\cite{Wang2011}.

The simplest method to measure the distance between two given genes is to detect whether their corresponding proteins are connected directly in the initial network or not \cite{Linghu2009,wu2008network}. In direct link (DL) method we score candidate genes according to the condition that they have a direct link $E_{i,j}$ to seed genes as shown in Equation \ref{equ-DL}.

\begin{equation}\label{equ-DL}
S_i =\left\{ \begin{array}{l}
\sum _{j \in S_{set}} W_{i,j}, \quad if \exists E_{i,j}\\
0, \quad \quad \quad \quad \quad Otherwise
\end{array} \right.\
\end{equation}

Dijkstra's shortest path is used to score candidate gene that have not direct link with seed genes but are involved in the same biological pathway \cite{wu2008network}.

Considering that local distance  measure do not have the ability to capture the overall network topology, we applied random walk as a global distance measure to tackle this issue.
Random walk with restarts (RWR) has a wide usage in disease gene prioritization in FLN and PPI network\cite{kohler2008walking,lei2019random, valdeolivas2018random, navlakha2010power}. The random walk method on network is defined as simulation of a random walk on the network to compute the closeness between two genes by capturing the global topology of the network. The RWR in disease gene prioritization task is defined as follows: an iterative walker starts at one of the genes in seed set $S_{set}$, then at each step moves to a randomly chosen neighbor $n\in N$ of the current gene. In comparison with simple random walk, the RWR can restart at one of the genes in the seed set $S_{set}$ every step with the probability of $r$. Formally, the RWR is defined as:

\begin{equation}\label{equ-RWR}
R_t = (1-r)WR_{t-1}+rR_{_0}
\end{equation}
where $W$ is the column-normalized adjacency matrix of the $FLN$ network, $R_t$ is the probability vector being at the supposed node at iteration $t$, and $R_0$ is the probability being restart at each seed node, which is computed as follows:

\begin{equation}\label{equ-RWRW}
W{i,j} =\left\{ \begin{array}{l}
 w_{i,j}/W(j), \quad if \exists E_{i,j}\\
0, \quad \quad \quad \quad \quad Otherwise
\end{array} \right.\
\end{equation}

\begin{equation}\label{equ-RWRP}
R_0=\left\{ \begin{array}{l}
\rho(s_{_0},S_{set})/\sum _{s \in S_{set}} \rho(s,S_{set}), \quad if s_{_0} \exists S_{set}\\
0, \quad \quad \quad \quad \quad \quad \quad \quad \quad \quad \quad Otherwise
\end{array} \right.\
\end{equation}
where $\rho(s_{_0},S_{set})$ is a function to calculate probability of restarting at $s_{_0} \in S $ that is defined as the degree of association between mentioned node $s_{_0}$ and the other seed genes.

\section{Results and Discussions}
\label{sec:results}
%In this section first the evaluation measurements, which are used to evaluate the efficiency of the proposed method, are introduced. Then we describe the dataset, that gold standards are created by changing some parameters.
%Finally, the effect of the different steps of the S-FLN approach on the gold standard datasets is investigated and results are presented.

\subsection{Experimental setup}
\label{sec:esetting}
We use the Leave-One-Out cross validation to evaluate the performance of the proposed methods in terms of accurately prioritizing candidate disease genes. To do this, for each gene in seed set the experiment manner is execute as follow:
\begin{itemize}
	\item The first gene is removed from the seed set, where this gene is called the target gene
	 that is considered as target candidate gene $c_t \in C_{set}$, in this iteration of experiment. The other known disease genes remain as the seed set $S_{set}$.
	\item We apply the artificial linkage interval approach to generate the new candidate set including  $c_t$ and the set of the nearest 99 genes around that in terms of genomic distance. 
	\item We used the aforementioned distance methods to prioritize the new candidate set described in second step, and the assigned rank of $c_t$.
\end{itemize}

The proposed framework will be evaluated in two different aspects:  1)Disease-centric evaluation, in which the performance of the framework is evaluated for each individual disease. 2)Gene-centric assessment, in which treats all seed genes of different disease as target candidate gene $c_t$ to asses the the performance of the framework. 

\subsection{Dataset}
\label{sec:dataset}
%We have used different way in order to create the gold standards.
We employed different datasets for performance evaluation of the proposed approach.
At first, three datasets including Uniprot Consortium \cite{Boutet2016}, Gene ontology data of human gene annotation (released in October 2013) \cite{gene2015gene}, and %\footnote[1]{http://geneontology.org/}
NCBI database \cite{geer2009ncbi} are used to obtain the proteins' IDs, GO annotations, and amino-acid sequences, respectively. 
Afterwards, all proteins in the sets are weighted via GO semantic similarity measure to obtain GSP and GSN sets.
Finally, a sequenced-based FLN is constructed based on these GSP and GSN sets. The constructed S-FLN include, 17,442 proteins and 889,307 functional linkages, that is used as initial network in disease gene prioritization task.
Nodes and edges of S-FLN represent the genes and their functionality weights, respectively. The proposed S-FLN has an increased accuracy and coverage compared to the integration-based methods.

In the following, we also use two different versions of disease gene set to compare the proposed method with some other well-known methods in similar scope. In the first data set ($DG_{set1}$), we extract 1,025 known disease genes from the OMIM database\cite{hamosh2005online,Linghu2013}, then these diseases are categorized into 110 disease gene sets, that each one contains at least 5 genes and on average 11 genes. In the second dataset ($DG_{set2}$) \cite{kohler2008walking} also, the 110 disease families include 783 genes, that are maximum contained 41 and minimum 3 genes, are extracted from the OMIM database\cite{hamosh2002online}.

\subsection{Evaluation Metrics}
In order to evaluate the performance of the S-FLN and different aspect of that, we use the various metrics, following evaluation metrics:

\begin{itemize}
	\item \textbf{Area Under Receiver Operating Characteristic (AUROC):} ROC plots the true-positive rate (TPR) versus the false-positive rate (FPR) at different rank cutoffs. The ROC values can be interpreted as fraction of the number of the seed genes above the rank cutoff versus the number of the seed genes below the rank cutoff:

\begin{equation}\label{equ-specificity}
1-specificity = \frac{{FP}}{{TN + FP}}
\end{equation}

\begin{equation}\label{equ-sensitivity}
sensitivity = \frac{{TP}}{{TP + FN}}
\end{equation}
where TP (true positive) is the number of seed genes above
the rank cutoff, FP (false positive) is the number of non-seed genes above rank the cutoff, FN (false negative) is the number of seed genes below the rank cutoff, and TN (true negative) is the number of non-seed genes below the rank cutoff.

AUCROC is not sufficient in this application, since there is only one target seed gene $c_t$ in each experiment iteration. Therefore, for this purpose we apply other evaluation metrics.

\item \textbf{Average rank:}
This measure shows the average rank of the $c_t$ among all candidate genes set $C_{set}$, that computed  in a 10-fold iteration. Clearly, lower value of average rank indicates better performance.
\item \textbf{Top n\%:} The top n\% shows the percentage of seed genes that are ranked below the thresholds of $n$  among all candidate gene set $C_{seed}$. We use of Top 1\% and 5\% to evaluation S-FLN.
\end{itemize}

\subsection{Prioritization Accuracy Comparison}

\begin{figure}[t]
	\centering
	\includegraphics[width=1\textwidth]{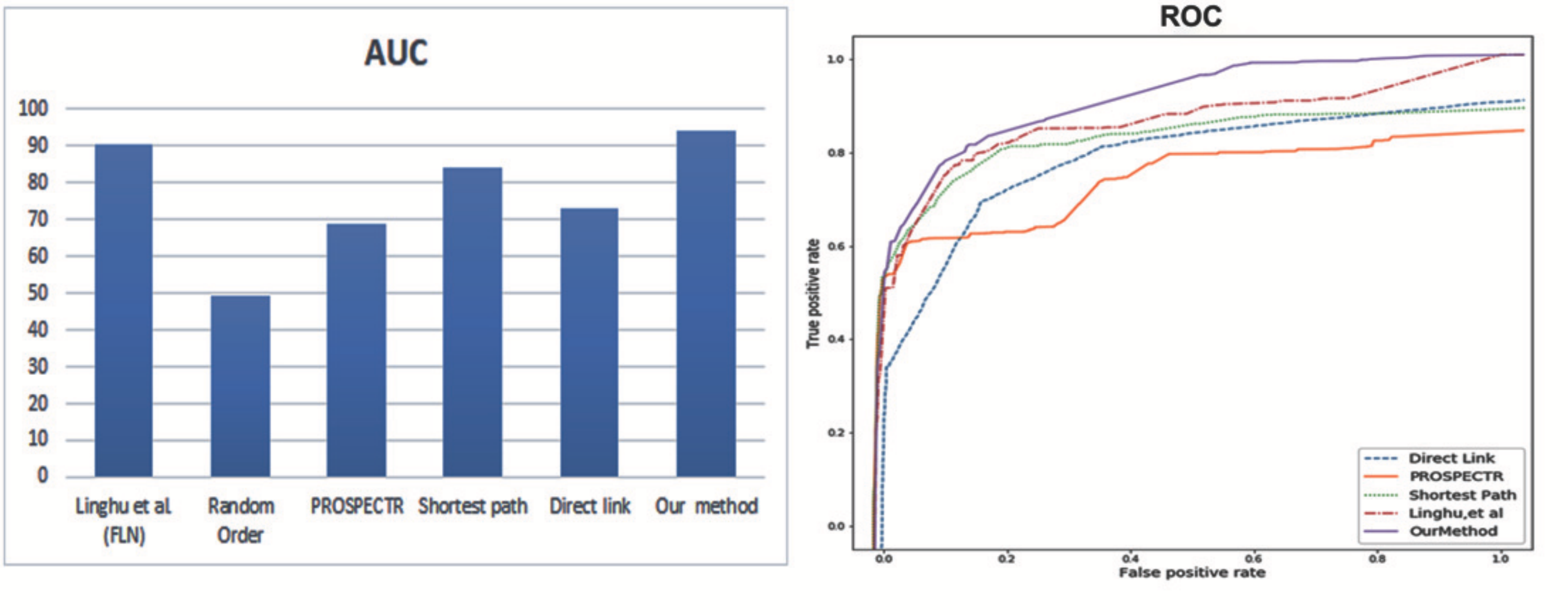}
	\caption{Comparison the S-FLN with state-of-the-art methods in disease gene prioritization in terms of the AUC values inside of ROC plots. The proposed S-FLN shows better result.}
	\label{fgr:globalAUC}
\end{figure}

In order to show the advantages of the proposed framework in disease gene prioritization application and to compare with the state-of-the-art prioritization methods, ($DG_{set2}$) of OMIM dataset has been used. We compare the result of the proposed framework based on Leave-One-Out cross-validation with the reported results by \cite{oti2006predicting,Linghu2009,Linghu2013,adie2006suspects} and two standard distance measure methods in the same data set. 

As shown in Figure \ref{fgr:globalAUC}, the AUC obtained by the proposed framework is 0.943 as the best performance record.
The PROSPECTR \cite{adie2006suspects} method shows the worst performance.
Because it does not consider the biological network information for prioritization task. The other compared methods are known as network-based methods. These methods are based on the assumption that genes related to a disease have interaction with each other.  
In the network-based methods, Direct Link and Shortest Path methods show moderate performances with a functional enriched PPI network\cite{kohler2008walking}. 
The obtained result of \cite{Linghu2013,Linghu2009} and the proposed S-FLN demonstrate that a functional linkage network is a valuable source to apply as initial network in prioritization task. 
The most important disadvantage of FLNs is the existence of noise in them, i.e. there is missing linkage or false linkage in constructed networks. However, the proposed S-FLN has overcome this problem, with the consideration of protein sequence information as a comprehensive information, which finally improves the result of disease gene prioritization.

\subsection{Analyzing the S-FLN}
In this subsection, different stages of S-FLN in prioritization process are discussed.
%\begin{figure*}[h]
% \centering
% \includegraphics[width=\textwidth=13]{compare-dataAB}
% \caption{An image from the \textit{Physical Chemistry Chemical Physics} cover gallery, set as a two-column figure.}
% \label{fgr:PCA-plot}
%\end{figure*}

\begin{figure}[t]
	\centering
	\includegraphics[width=1\textwidth]{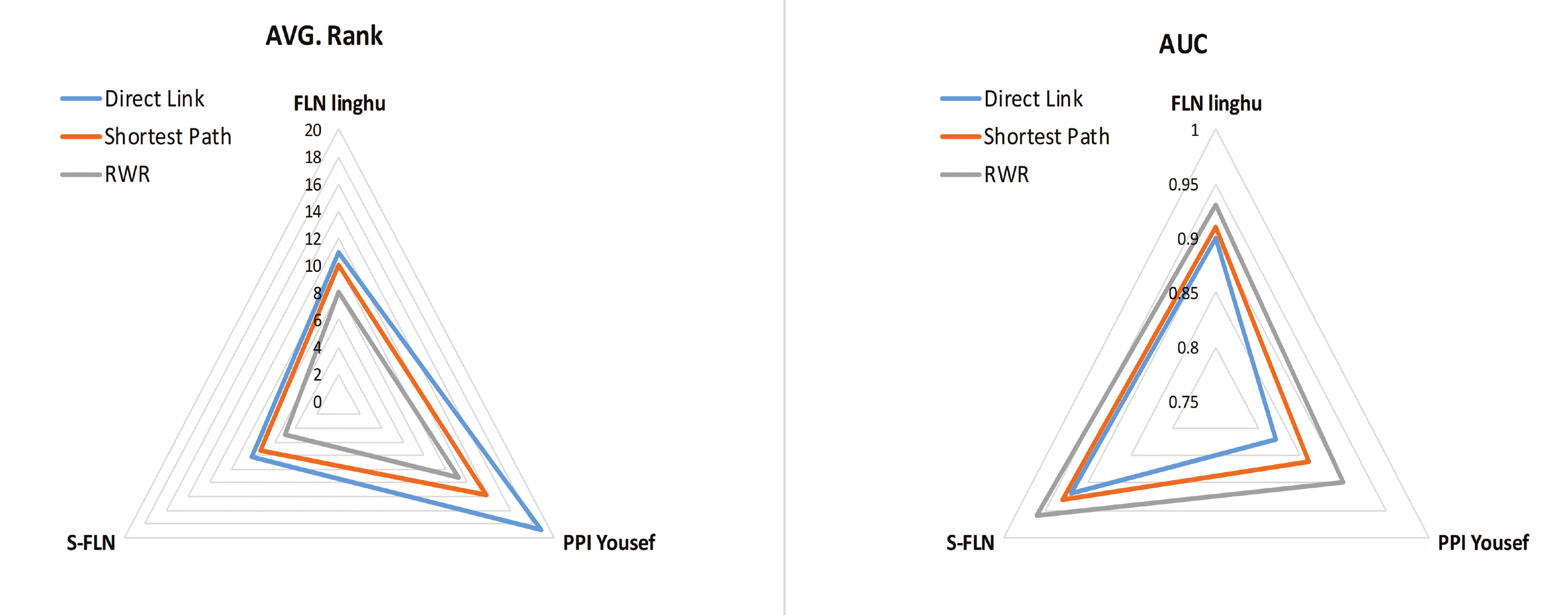}
	\caption{Comparison of Average-Rank and AUC evaluation metrics in the three initial networks by three different distance measures.}
	\label{fgr:AvgRankAUC}
\end{figure}

%\subsubsection{Impact of semantic similarity weight on performance}
\textbf{Initial network}:
As mentioned in section \ref{sec:dataset}, we constructed a FLN including 17,442 proteins and 889,307 functional linkages to use as the initial network in disease gene prioritization task. The S-FLN is compared with two biological networks, which include a PPI and a FLN network, whereas their results become competitive with S-FLN as initial network. To do this, we employ three introduced distance measures with the ($DG_{set2}$) omim disease set on these networks. The obtained results are evaluated by two evaluation metrics including AUC and Average Rank.

As it can be observed from Figure \ref{fgr:AvgRankAUC}, the best performance is obtained by the proposed framework, that is the result of  employing physicochemical properties of amino-acid by different features descriptors as well as developing an ensemble based learning algorithm. 
This experiment shows that the primary data sources, which are used to construct the initial network, have important role in the results of the disease gene prioritization. That is why the proposed framework, which is based on S-FLN, shows better results compared with the integration-based FLN\cite{Linghu2013}, where protein sequence data are used to represent functional similarity of gene products as a high coverage, high quality, and comprehensive data source.

According to the fact that disease genes generally are in a collaboration to generate a disease pathway instead of having physically interacting, the DL distance measure fails to prioritize candidate genes that have no direct link to the seed genes. It is observed in Figure \ref{fgr:AvgRankAUC}, that difference between DL measure and other measures in the PPI networks is more than the other two FLN networks.

In Table\ref{tbl:compare-other}, the mentioned initial networks are evaluated with more evaluation metrics including Average Rank, AUC, 1\%, and 5\% top genes. 
In this evaluation, $S-FLN$ and $S-FLN (DG{set2})$ successfully rank 36.5\% and 29.5\% disease genes as rank 1, respectively. Whereas, FLN\cite{Linghu2013} and \cite{yousef2013novel} only rank 34.2\% and 28.25\% disease genes as rank 1, respectively.
There is a new setup, where we made use of ($DG_{set2}$) omim disease set inside S-FLN to evaluate impact of disease set on prioritization task.  
As show in Table\ref{tbl:compare-other}, the AUC is decreased about 2\% when $DG_{set2}$ is used, which means that this dataset includes fewer disease genes.
It should be noticed that the non-overlapping genes of $DG_{set1}$ has been used as candidate set $C_{set}$ in new setup $S-FLN$ $RWR(DG_{set2})$, which confirms S-FLN still has better performance than that in the other networks independent of the disease set.

\begin{table}[t]
	\small
	\caption{\ Comparing the S-FLN with other biological networks in terms of the Average-Rank, AUC, Top 1\%, and Top 5\%. In all networks, the Random Walk with Restart is used as distance measure to obtain fair evaluation setup.}
	\label{tbl:compare-other}
	%  \resizebox{0.45\textwidth}{!}{
	\begin{tabular*}{0.74\textwidth}{@{\extracolsep{\fill}}c|cccc}
		\hline
		&\multirow{2}{*}{ \shortstack{FLN\\RWR\cite{Linghu2013}}}  &\multirow{2}{*}{\shortstack{PPI\\RWR\cite{yousef2013novel}}} &\multirow{2}{*}{ \shortstack{S-FLN\\RWR($DG_{set2}$)}} &\multirow{2}{*}{\bf \shortstack{S-FLN\\RWR}} \\
		&&&&\\
		\hline
		Avg. Rank & 8.03 & 11.1 &7.1& 5\\
		AUC  & 0.93 & 0.90 & 0.94 &0.96\\
		Top 1\% & 34.2 & 28.25 & 29.5& 36.5\\
		Top 5\% & 62.3 & 69.3 & 70.8 & 83\\
		\hline
	\end{tabular*}%}
\end{table}

\begin{table}[h]
	\small
	\caption{\ Comparing the results of the S-FLN with that of the state-of-the-art methods in disease gene prioritization in terms of ranking new disease genes. In the experiments we used  $DG_{set1}$ as primary disease gene set and new introduced gene in $DG_{set2}$ as the test set $c_t$. }
	\label{tbl:NewGenes}
	%  \resizebox{0.45\textwidth}{!}{
	\setlength{\extrarowheight}{2pt}
	\begin{tabular*}{\textwidth}{c||c||ccccccc}
		\hline
		\multirow{2}{*}{\bf \shortstack{Disease \\Family}} & \multirow{2}{*}{\bf \shortstack{Gene \\ Name}} &\multirow{2}{*}{ \shortstack{FLN\\RWR\cite{kohler2008walking}}} & \multirow{2}{*}{\shortstack{FLN\\SP\cite{kohler2008walking}}} & \multirow{2}{*}{ \shortstack{FLN\\DL\cite{kohler2008walking}}}& \multirow{2}{*}{ \shortstack{FLN\\DL\cite{Linghu2013}}}  & \multirow{2}{*}{\shortstack{PPI\\web\cite{aerts2006gene}}} & \multirow{2}{*}{\shortstack{PROSPECTR\\SB\cite{adie2006suspects}}} &\multirow{2}{*}{\bf \shortstack{S-FLN\\RWR}}\\
		&&&&&&&& \\
		\hline
		\hline
		Nephronophthisis & GLIS237 & 100 & 100 & 100 & 100 & 43 & \textbf{3} & 21\\
		Retinitis Pig. & TOPORS39 & 23 & 20 & 100 & 23 & 69 & 56 & \textbf{16}\\
		Retinitis Pig. & NR2E340 & 2 & 18 & 100 & \textbf{1} & 2 & \textbf{1} & \textbf{1}\\
		Noonan Syndrome & RAF141 & \textbf{1} & 4 & 4 & \textbf{1} & 3 & 42 & \textbf{1}\\
		Brachydactyly & NOG42 & \textbf{1} & \textbf{1} & \textbf{1} & \textbf{1} & 5 & 34 & \textbf{1}\\
		CMT4H & FGD443 & 13 & 27 & 100 & 4 & \textbf{2} & 9 & \textbf{2}\\
		\hline
	\end{tabular*}%}
\end{table}

\textbf{Disease-centeric evaluation}:
According to various setups taken in other researches (different initial network, different diseases set, and different distance measures), 
we evaluate the performance of the S-FLN in disease gene prioritization by simulating the real-life search for new disease genes that are not involved in known disease datasets. 
Therefore, in this experiment we compare the S-FLN with state-of-the-art methods in disease gene prioritization in terms of ranking new disease genes, where we used $DG_{set1}$ as primary disease gene set (train set). We also used  six new disease genes introduced in $DG_{set2}$ that do not overlap with $DG_{set1}$ as test set $c_t$.
As shown in Table \ref{tbl:NewGenes}, FLN-based methods are unable to prioritize the gene $GLIS237$ of $Nephronophthisis$ disease family. In contrast, PROSPECTR as a sequence analysis method has the best performance on the $GLIS237$ gene. The S-FLN obtains the second performance rate on the $GLIS237$ gene. It is because the S-FLN employs physicochemical properties of amino-acids to construct FLN. In other cases, the S-FLN has obtained the best performance.  

\begin{table}[h]
	\small
	\caption{\  Top 10 ranked disease genes for 3 case studies diseases with RWR on the proposed S-FLN and $DG_{set1}$.}
	\label{tbl:10top3disease}
	%  \resizebox{0.45\textwidth}{!}{
	\setlength{\extrarowheight}{10pt}
	\begin{tabular*}{0.65\textwidth}{@{\extracolsep{\fill}} c|ccc}
		\hline 
		\multirow{2}{*}{\bf \shortstack{Disease \\ Family}} & \multirow{2}{*}{\bf \shortstack{ Breast \\cancer}} & \multirow{2}{*}{\bf \shortstack{Alzheimer \\disease}} & \multirow{2}{*}{\bf \shortstack{Diabetes \\mellitus}} \\
		&&&\\
		%\addlinespace[12pt] 
		\hline
		\parbox[t]{2mm}{\multirow{10}{*}{\rotatebox[origin=c]{90}{Top-10 predictions genes}}}
		& MSH2* & \textbf{APP} & TCF1 \\
		&\textbf{BRCA1} & \textbf{MAPT} & \textbf{TCF7L2} \\
		&\textbf{PTEN} & \textbf{PSEN2} & \textbf{TCF2} \\ 
		&MSH6* & TREM2* & \textbf{INS} \\
		&NBN & \textbf{PSEN1} & TCF7*\\
		&\textbf{PIK3CA} & \textbf{APOE} & CTNNB1*\\
		&\textbf{BRCA2} & \textbf{LRP1} & \textbf{AKT2} \\
		&\textbf{TP53} & TYROBP* & THBS2* \\
		&MLH1 & HD & APC \\
		&BCL2* & CST3 & CTNNA1* \\
		
		\hline
	\end{tabular*}%}
\end{table}

%\subsubsection{Impact of train set on the performance}
Moreover, we have selected three disease families as case studies to evaluate the S-FLN by disease-center evaluation approach. Table \ref{tbl:10top3disease} shows 10  disease genes which obtained ten top ranks for three disease families by RWR distance measure on S-FLN. These disease families are $Breast\ cancer$, $Alzheimer\ disease$, and $Diabetes\ mellitus$.

The results of Table \ref{tbl:10top3disease} indicate the performance of S-FLN in prioritizing known disease genes and predicting new susceptible candidate genes, where the $Breast Cancer$ is considered as the first case study. Among 10 genes ranked by the S-FLN 5 genes are shown in Bold face that are, as reported, known disease gene in OMIM(MIM 114480). Additionally, the other 3 genes which are ranked and show by $*$ symbol, are introduced as new breast cancer susceptible genes by literature including PMID:29345684, pubmed:28779004, and PMID: 27721874.
These results are shown for $Alzheimer Disease$(MIM 104300) and $Diabetes Mellitus$ (MIM 125853) in the same style, where new susceptible genes by literature for these disease are reported in (PMID:23150934, PMID:25052481) and (PMID:22247771 PMID:27613243 PMID:27446820), respectively.

\section{Concluding Remarks}
In this paper, the problem of disease gene prioritization has been addressed as a network-based problem. Therefore, we proposed a hierarchical two step approach. The main important step in network-based disease gene prioritization approaches is applying a comprehensive initial network. Commonly, more than one data source is employed to construct the FLN as an applicable initial network. However, using integration approach to construct the FLN suffers from appearing some missing or invalid information which causes an undesirable performance in the constructing FLN and more in prioritization accuracy. In order to tackle this problem, a sequence-based approach is proposed to construct FLN as initial network, which is as the first step in the proposed method. 
In the second step, we develop different distance measures to compute network properties and distance between candidate genes and known disease gene as a score to rank candidate genes.  The proposed approach provides some interesting achievements. 
Thanks to the FLN which is used instead of physical interaction (PPI),  more disease genes are corporate in a pathway. Besides, in proposed approach we construct an FLN  using physicochemical properties of amino-acids, that makes the initial network able to model relations among disease genes as well. 
Furthermore, as the sequence of proteins is a more comprehensive and accurate data source, an applicable and reliable FLN is obtained.  The issue of missing values and invalid information is addressed as well. 
In the proposed approach, we have considered the single type link prediction task to construct the FLN. However, a typical FLN naturally includes various types of links and proteins dependency. In this regard, constructing the FLN according to a heterogeneous network model and consequently prioritizing disease genes based on it has been considered as our future plan.
\bibliography{disease} %You need to replace "rsc" on this line with the name of your .bib file

\begin{thebibliography}{10}
\expandafter\ifx\csname url\endcsname\relax
  \def\url#1{\texttt{#1}}\fi
\expandafter\ifx\csname urlprefix\endcsname\relax\def\urlprefix{URL }\fi
\expandafter\ifx\csname href\endcsname\relax
  \def\href#1#2{#2} \def\path#1{#1}\fi

\bibitem{tenesa2013heritability}
A.~Tenesa, C.~S. Haley, The heritability of human disease: estimation, uses and
  abuses, Nature Reviews Genetics 14~(2) (2013) 139.

\bibitem{lei2019predicting}
X.~Lei, Y.~Zhang, Predicting disease-genes based on network information loss
  and protein complexes in heterogeneous network, Information Sciences 479
  (2019) 386--400.

\bibitem{risch1996future}
N.~Risch, K.~Merikangas, The future of genetic studies of complex human
  diseases, Science 273~(5281) (1996) 1516--1517.

\bibitem{Vandamexpression}
S.~van Dam, U.~V{\~o}sa, A.~van~der Graaf, L.~Franke, J.~P. de~Magalh{\~a}es,
  Gene co-expression analysis for functional classification and gene disease
  predictions, Briefings in Bioinformatics (2017) bbw139\href
  {http://dx.doi.org/10.1093/bib/bbw139} {\path{doi:10.1093/bib/bbw139}}.

\bibitem{Tiffin}
N.~Tiffin, J.~F. Kelso, A.~R. Powell, H.~Pan, V.~B. Bajic, W.~A. Hide,
  Integration of text- and data-mining using ontologies successfully selects
  disease gene candidates, Nucleic Acids Research 33~(5) (2005) 1544--1552.
\newblock \href {http://dx.doi.org/10.1093/nar/gki296}
  {\path{doi:10.1093/nar/gki296}}.

\bibitem{eguchi2018integrative}
R.~Eguchi, M.~B. Karim, P.~Hu, T.~Sato, N.~Ono, S.~Kanaya, M.~Altaf-Ul-Amin, An
  integrative network-based approach to identify novel disease genes and
  pathways: a case study in the context of inflammatory bowel disease, BMC
  bioinformatics 19~(1) (2018) 264.

\bibitem{aerts2006gene}
S.~Aerts, D.~Lambrechts, S.~Maity, P.~Van~Loo, B.~Coessens, F.~De~Smet, L.-C.
  Tranchevent, B.~De~Moor, P.~Marynen, B.~Hassan, et~al., Gene prioritization
  through genomic data fusion, Nature biotechnology 24~(5) (2006) 537.

\bibitem{wang2011integration}
J.~Wang, G.~Chen, M.~Li, Y.~Pan, Integration of breast cancer gene signatures
  based on graph centrality, BMC systems biology 5~(3) (2011) S10.

\bibitem{liu2015prioritization}
B.~Liu, M.~Jin, P.~Zeng, Prioritization of candidate disease genes by combining
  topological similarity and semantic similarity, Journal of biomedical
  informatics 57 (2015) 1--5.

\bibitem{findlay2018guilt}
R.~Findlay, O.~Thompson, B.~A. Ziganshin, J.~A. Elefteriades, Guilt by
  association: Paradigm for detection of silent aortic aneurysms, in: New
  Approaches to Aortic Diseases from Valve to Abdominal Bifurcation, Elsevier,
  2018, pp. 107--118.

\bibitem{barabasi2011network}
A.-L. Barab{\'a}si, N.~Gulbahce, J.~Loscalzo, Network medicine: a network-based
  approach to human disease, Nature Reviews Genetics 12~(1) (2011) 56--68.

\bibitem{he2018evolutionary}
T.~He, K.~C. Chan, Evolutionary graph clustering for protein complex
  identification, IEEE/ACM transactions on computational biology and
  bioinformatics 15~(3) (2018) 892--904.

\bibitem{bbz017}
M.~Li, W.~Li, R.~Zheng, X.~Li, M.~Zeng, {Network-based methods for predicting
  essential genes or proteins: a survey}, Briefings in bioinformatics (2019).

\bibitem{herzog2018detection}
M.~Herzog, F.~Puddu, J.~Coates, N.~Geisler, J.~V. Forment, S.~P. Jackson,
  Detection of functional protein domains by unbiased genome-wide forward
  genetic screening, Scientific reports 8~(1) (2018) 6161.

\bibitem{schaefer2012evolution}
M.~H. Schaefer, E.~E. Wanker, M.~A. Andrade-Navarro, Evolution and function of
  cag/polyglutamine repeats in protein--protein interaction networks, Nucleic
  acids research 40~(10) (2012) 4273--4287.

\bibitem{wong2015imp}
A.~K. Wong, A.~Krishnan, V.~Yao, A.~Tadych, O.~G. Troyanskaya, Imp 2.0: a
  multi-species functional genomics portal for integration, visualization and
  prediction of protein functions and networks, Nucleic acids research 43~(W1)
  (2015) W128--W133.

\bibitem{oti2006predicting}
M.~Oti, B.~Snel, M.~A. Huynen, H.~G. Brunner, Predicting disease genes using
  protein--protein interactions, Journal of medical genetics 43~(8) (2006)
  691--698.

\bibitem{hsu2011prioritizing}
C.-L. Hsu, Y.-H. Huang, C.-T. Hsu, U.-C. Yang, Prioritizing disease candidate
  genes by a gene interconnectedness-based approach, in: BMC genomics, Vol.~12,
  BioMed Central, 2011, p. S25.

\bibitem{zhu2012vertex}
C.~Zhu, A.~Kushwaha, K.~Berman, A.~G. Jegga, A vertex similarity-based
  framework to discover and rank orphan disease-related genes, BMC systems
  biology 6~(3) (2012) S8.

\bibitem{li2014prioritization}
M.~Li, Q.~Li, G.~U. Ganegoda, J.~Wang, F.~Wu, Y.~Pan, Prioritization of orphan
  disease-causing genes using topological feature and go similarity between
  proteins in interaction networks, Science China Life Sciences 57~(11) (2014)
  1064--1071.

\bibitem{Wang2011}
X.~Wang, N.~Gulbahce, H.~Yu,
  \href{http://www.ncbi.nlm.nih.gov/pubmed/21764832}{{Network-based methods for
  human disease gene prediction.}}, Briefings in functional genomics 10~(5)
  (2011) 280--93.
\newblock \href {http://dx.doi.org/10.1093/bfgp/elr024}
  {\path{doi:10.1093/bfgp/elr024}}.
\newline\urlprefix\url{http://www.ncbi.nlm.nih.gov/pubmed/21764832}

\bibitem{manimaran2009prediction}
P.~Manimaran, S.~R. Hegde, S.~C. Mande, Prediction of conditional gene
  essentiality through graph theoretical analysis of genome-wide functional
  linkages, Molecular Biosystems 5~(12) (2009) 1936--1942.

\bibitem{apolloni2011learning}
B.~Apolloni, et~al., Learning functional linkage networks with a cost-sensitive
  approach, in: Neural Nets WIRN10: Proceedings of the 20th Italian Workshop on
  Neural Nets, Vol. 226, IOS Press, 2011, p.~52.

\bibitem{jalilvand2018s}
A.~Jalilvand, B.~Akbari, F.~Z. Mirakabad, S-fln: A sequence-based hierarchical
  approach for functional linkage network construction, Journal of theoretical
  biology 437 (2018) 149--162.

\bibitem{you2010using}
Z.-H. You, Y.-K. Lei, J.~Gui, D.-S. Huang, X.~Zhou, Using manifold embedding
  for assessing and predicting protein interactions from high-throughput
  experimental data, Bioinformatics 26~(21) (2010) 2744--2751.

\bibitem{lei2012assessing}
Y.-K. Lei, Z.-H. You, Z.~Ji, L.~Zhu, D.-S. Huang, Assessing and predicting
  protein interactions by combining manifold embedding with multiple
  information integration, BMC bioinformatics 13~(7) (2012) 1.

\bibitem{Franke2006}
L.~Franke, H.~van Bakel, L.~Fokkens, E.~de~Jong, M.~Egmont-Petersen,
  C.~Wijmenga,
  \href{http://pdn.sciencedirect.com/science?{\_}ob=MiamiImageURL{\&}{\_}cid=276895{\&}{\_}user=1517318{\&}{\_}pii=S0002929707639226{\&}{\_}check=y{\&}{\_}origin=article{\&}{\_}zone=toolbar{\&}{\_}coverDate=30-Jun-2006{\&}view=c{\&}originContentFamily=serial{\&}wchp=dGLzVlk-zSkzk{\&}md5=6cad6aa11685150b3f3c23cfff694942}{{Reconstruction
  of a functional human gene network, with an application for prioritizing
  positional candidate genes}}, Am J Hum Genet 78~(June) (2006) 1011----1025.
\newline\urlprefix\url{http://pdn.sciencedirect.com/science?{\_}ob=MiamiImageURL{\&}{\_}cid=276895{\&}{\_}user=1517318{\&}{\_}pii=S0002929707639226{\&}{\_}check=y{\&}{\_}origin=article{\&}{\_}zone=toolbar{\&}{\_}coverDate=30-Jun-2006{\&}view=c{\&}originContentFamily=serial{\&}wchp=dGLzVlk-zSkzk{\&}md5=6cad6aa11685150b3f3c23cfff694942}

\bibitem{Linghu2009}
B.~Linghu, E.~S. Snitkin, Z.~Hu, Y.~Xia, C.~DeLisi, Genome-wide prioritization
  of disease genes and identification of disease-disease associations from an
  integrated human functional linkage network, Genome biology 10~(9) (2009)
  R91.

\bibitem{Wang2014}
J.~Wang, J.~Yang, S.~Mao, X.~Chai, Y.~Hu, X.~Hou, Y.~Tang, C.~Bi, X.~Li,
  \href{http://dx.plos.org/10.1371/journal.pone.0111187}{{MitProNet: A
  Knowledgebase and Analysis Platform of Proteome, Interactome and Diseases for
  Mammalian Mitochondria}}, PLoS ONE 9~(10) (2014) e111187.
\newblock \href {http://dx.doi.org/10.1371/journal.pone.0111187}
  {\path{doi:10.1371/journal.pone.0111187}}.
\newline\urlprefix\url{http://dx.plos.org/10.1371/journal.pone.0111187}

\bibitem{Wu2010}
M.~Wu, X.~Li, H.~N. Chua, C.-K. Kwoh, S.-K. Ng, Integrating diverse biological
  and computational sources for reliable protein-protein interactions, BMC
  bioinformatics 11~(7) (2010) S8.

\bibitem{kohler2008walking}
S.~K{\"o}hler, S.~Bauer, D.~Horn, P.~N. Robinson, Walking the interactome for
  prioritization of candidate disease genes, The American Journal of Human
  Genetics 82~(4) (2008) 949--958.

\bibitem{Linghu2008}
B.~Linghu, E.~S. Snitkin, D.~T. Holloway, A.~M. Gustafson, Y.~Xia, C.~DeLisi,
  {High-precision high-coverage functional inference from integrated data
  sources.}, BMC bioinformatics 9 (2008) 119.
\newblock \href {http://dx.doi.org/10.1186/1471-2105-9-119}
  {\path{doi:10.1186/1471-2105-9-119}}.

\bibitem{Linghu2013}
B.~Linghu, E.~A. Franzosa, Y.~Xia, {Construction of functional linkage gene
  networks by data integration}, in: Data Mining for Systems Biology, Springer,
  2013, pp. 215--232.

\bibitem{singh2018evolutionary}
D.~Singh, P.~Singh, D.~S. Sisodia, Evolutionary based ensemble framework for
  realizing transfer learning in hiv-1 protease cleavage sites prediction,
  Applied Intelligence (2018) 1--23.

\bibitem{shen2007predicting}
J.~Shen, J.~Zhang, X.~Luo, W.~Zhu, K.~Yu, K.~Chen, Y.~Li, H.~Jiang, Predicting
  protein--protein interactions based only on sequences information,
  Proceedings of the National Academy of Sciences 104~(11) (2007) 4337--4341.

\bibitem{guo2008using}
Y.~Guo, L.~Yu, Z.~Wen, M.~Li, Using support vector machine combined with auto
  covariance to predict protein--protein interactions from protein sequences,
  Nucleic acids research 36~(9) (2008) 3025--3030.

\bibitem{xia2010predicting}
J.-F. Xia, X.-M. Zhao, D.-S. Huang, Predicting protein--protein interactions
  from protein sequences using meta predictor, Amino Acids 39~(5) (2010)
  1595--1599.

\bibitem{yousef2013novel}
A.~Yousef, N.~M. Charkari, A novel method based on new adaptive lvq neural
  network for predicting protein--protein interactions from protein sequences,
  Journal of theoretical biology 336 (2013) 231--239.

\bibitem{you2013prediction}
Z.-H. You, Y.-K. Lei, L.~Zhu, J.~Xia, B.~Wang, Prediction of protein-protein
  interactions from amino acid sequences with ensemble extreme learning
  machines and principal component analysis, BMC bioinformatics 14~(8) (2013)
  1.

\bibitem{mei2014adaboost}
S.~Mei, H.~Zhu, Adaboost based multi-instance transfer learning for predicting
  proteome-wide interactions between salmonella and human proteins, PloS one
  9~(10) (2014) e110488.

\bibitem{borozan2015integrating}
I.~Borozan, S.~Watt, V.~Ferretti, Integrating alignment-based and
  alignment-free sequence similarity measures for biological sequence
  classification, Bioinformatics (2015) btv006.

\bibitem{Yushuang2016}
Y.~Li, T.~Song, J.~Yang, Y.~Zhang, J.~Yang, An alignment-free algorithm in
  comparing the similarity of protein sequences based on pseudo-markov
  transition probabilities among amino acids, PLOS ONE 11~(12) (2016) 1--14.

\bibitem{otu2003new}
H.~H. Otu, K.~Sayood, A new sequence distance measure for phylogenetic tree
  construction, Bioinformatics 19~(16) (2003) 2122--2130.

\bibitem{zhang2011adaptive}
Y.-N. Zhang, X.-Y. Pan, Y.~Huang, H.-B. Shen, Adaptive compressive learning for
  prediction of protein--protein interactions from primary sequence, Journal of
  theoretical biology 283~(1) (2011) 44--52.

\bibitem{yang2010prediction}
L.~Yang, J.-F. Xia, J.~Gui, Prediction of protein-protein interactions from
  protein sequence using local descriptors, Protein and peptide letters 17~(9)
  (2010) 1085--1090.

\bibitem{huang2016sequence}
Y.-A. Huang, Z.-H. You, X.~Chen, K.~Chan, X.~Luo, Sequence-based prediction of
  protein-protein interactions using weighted sparse representation model
  combined with global encoding, BMC bioinformatics 17~(1) (2016) 184.

\bibitem{sokal2006population}
R.~R. Sokal, B.~A. Thomson, Population structure inferred by local spatial
  autocorrelation: an example from an amerindian tribal population, American
  journal of physical anthropology 129~(1) (2006) 121--131.

\bibitem{xia2010sequence}
J.-F. Xia, K.~Han, D.-S. Huang, Sequence-based prediction of protein-protein
  interactions by means of rotation forest and autocorrelation descriptor,
  Protein and Peptide Letters 17~(1) (2010) 137--145.

\bibitem{feng2000prediction}
Z.-P. Feng, C.-T. Zhang, Prediction of membrane protein types based on the
  hydrophobic index of amino acids, Journal of protein chemistry 19~(4) (2000)
  269--275.

\bibitem{song2014measure}
X.~Song, L.~Li, P.~K. Srimani, P.~S. Yu, J.~Z. Wang, Measure the semantic
  similarity of go terms using aggregate information content, IEEE/ACM
  Transactions on Computational Biology and Bioinformatics (TCBB) 11~(3) (2014)
  468--476.

\bibitem{sweet1983correlation}
R.~M. Sweet, D.~Eisenberg, Correlation of sequence hydrophobicities measures
  similarity in three-dimensional protein structure, Journal of molecular
  biology 171~(4) (1983) 479--488.

\bibitem{hopp1981prediction}
T.~P. Hopp, K.~R. Woods, Prediction of protein antigenic determinants from
  amino acid sequences, Proceedings of the National Academy of Sciences 78~(6)
  (1981) 3824--3828.

\bibitem{grantham1974amino}
R.~Grantham, Amino acid difference formula to help explain protein evolution,
  Science 185~(4154) (1974) 862--864.

\bibitem{charton1982structural}
M.~Charton, B.~I. Charton, The structural dependence of amino acid
  hydrophobicity parameters, Journal of theoretical biology 99~(4) (1982)
  629--644.

\bibitem{eisenberg1986solvation}
D.~Eisenberg, A.~D. McLachlan, Solvation energy in protein folding and binding,
  Nature 319~(6050) (1986) 199--203.

\bibitem{fauchere1988amino}
J.-L. FAUCH{\`E}RE, M.~Charton, L.~B. Kier, A.~Verloop, V.~Pliska, Amino acid
  side chain parameters for correlation studies in biology and pharmacology,
  Chemical Biology \& Drug Design 32~(4) (1988) 269--278.

\bibitem{janin1979surface}
J.~Janin, Surface and inside volumes in globular proteins, Nature 277~(5696)
  (1979) 491--492.

\bibitem{prabhakaran1982shape}
M.~Prabhakaran, P.~Ponnuswamy, Shape and surface features of globular proteins,
  Macromolecules 15~(2) (1982) 314--320.

\bibitem{chothia1976nature}
C.~Chothia, The nature of the accessible and buried surfaces in proteins,
  Journal of molecular biology 105~(1) (1976) 1--12.

\bibitem{garel1973coefficients}
J.~P. Garel, D.~Filliol, P.~Mandel, Coefficients de partage daminoacides,
  nucleobases, nucleosides et nucleotides dans un systeme solvant salin,
  Journal of Chromatography A 78~(2) (1973) 381--391.

\bibitem{hutchens1970heat}
J.~O. Hutchens, Heat capacities, absolute entropies, and entropies of formation
  of amino acids and related compounds, Handbook of biochemistry.

\bibitem{galar2012review}
M.~Galar, A.~Fernandez, E.~Barrenechea, H.~Bustince, F.~Herrera, A review on
  ensembles for the class imbalance problem: bagging-, boosting-, and
  hybrid-based approaches, IEEE Transactions on Systems, Man, and Cybernetics,
  Part C (Applications and Reviews) 42~(4) (2012) 463--484.

\bibitem{wu2008network}
X.~Wu, R.~Jiang, M.~Q. Zhang, S.~Li, Network-based global inference of human
  disease genes, Molecular systems biology 4~(1) (2008) 189.

\bibitem{lei2019random}
X.~Lei, X.~Yang, H.~Fujita, Random walk based method to identify essential
  proteins by integrating network topology and biological characteristics,
  Knowledge-Based Systems.

\bibitem{valdeolivas2018random}
A.~Valdeolivas, L.~Tichit, C.~Navarro, S.~Perrin, G.~Odelin, N.~Levy, P.~Cau,
  E.~Remy, A.~Baudot, Random walk with restart on multiplex and heterogeneous
  biological networks, Bioinformatics 35~(3) (2018) 497--505.

\bibitem{navlakha2010power}
S.~Navlakha, C.~Kingsford, The power of protein interaction networks for
  associating genes with diseases, Bioinformatics 26~(8) (2010) 1057--1063.

\bibitem{Boutet2016}
E.~Boutet, D.~Lieberherr, M.~Tognolli, M.~Schneider, P.~Bansal, A.~J. Bridge,
  S.~Poux, L.~Bougueleret, I.~Xenarios, Uniprotkb/swiss-prot, the manually
  annotated section of the uniprot knowledgebase: how to use the entry view,
  Plant Bioinformatics: Methods and Protocols (2016) 23--54.

\bibitem{gene2015gene}
G.~O. Consortium, et~al., Gene ontology consortium: going forward, Nucleic
  acids research 43~(D1) (2015) D1049--D1056.

\bibitem{geer2009ncbi}
L.~Y. Geer, A.~Marchler-Bauer, R.~C. Geer, L.~Han, J.~He, S.~He, C.~Liu,
  W.~Shi, S.~H. Bryant, The ncbi biosystems database, Nucleic acids research
  (2009) gkp858.

\bibitem{hamosh2005online}
A.~Hamosh, A.~F. Scott, J.~S. Amberger, C.~A. Bocchini, V.~A. McKusick, Online
  mendelian inheritance in man (omim), a knowledgebase of human genes and
  genetic disorders, Nucleic acids research 33~(suppl\_1) (2005) D514--D517.

\bibitem{hamosh2002online}
A.~Hamosh, A.~F. Scott, J.~Amberger, C.~Bocchini, D.~Valle, V.~A. McKusick,
  Online mendelian inheritance in man (omim), a knowledgebase of human genes
  and genetic disorders, Nucleic acids research 30~(1) (2002) 52--55.

\bibitem{adie2006suspects}
E.~A. Adie, R.~R. Adams, K.~L. Evans, D.~J. Porteous, B.~S. Pickard, Suspects:
  enabling fast and effective prioritization of positional candidates,
  Bioinformatics 22~(6) (2006) 773--774.

\end{thebibliography}
\bibliographystyle{elsarticle-num} %the RSC's .bst file
%
%\begin{thebibliography}{00}
%\bibitem{OLAP}
%Loudcher S, Jakawat W, Morales EP, Favre C. "Combining OLAP and information networks for bibliographic data analysis: a survey." Scientometrics. 2015 May 1;103(2):471-87.
%\bibitem{LBSN}
%Miritello, Giovanna, Esteban Moro, Rubén Lara, Rocío Martínez-López, John Belchamber, Sam GB Roberts, and Robin IM Dunbar. "Time as a limited resource: Communication strategy in mobile phone networks." Social Networks 35, no. 1 (2013): 89-95.
%\end{thebibliography}
\end{document}